# Electric Axle and Wheel Module Driveline Concepts for Self-propelled Agricultural Machinery and Equipment Carriers

**Timo Oksanen, Karl Th. Renius**

*Technical University of Munich, Germany;*
*Chair of Agrimechatronics and*
*Munich Institute of Robotics and Machine Intelligence (MIRMI)*
*e-mail: timo.oksanen@tum.de, renius@tum.de*

**Abstract**: Direct electric drivelines without power-split open new design freedom for frame and suspension design, along with often lower energy losses. This paper focuses on self-propelled agricultural machinery (combine and forage harvesters, root crop harvesters), equipment carriers, propelled trailers and field robots. For a typical vehicle with four driven wheels, the electric motors can be packaged as two axle modules or four wheel modules, both defined herein as self-contained mechatronic units with integrated power electronics, distributed control intelligence and steering. Axle module and wheel module concepts are compared in detail against engineering requirements including loads, efficiency, steerability, controllability, braking, suspension, structural load support, asymmetric wheel loading and manufacturing cost. The wheel module offers maximum design freedom, redundancy and controllability, while the axle module provides lower cost, structural rigidity, automatic load sharing through the differential and the ability to be used in existing vehicle structures. Both concepts are defined such that distributed control intelligence and steering are integral to each unit, requiring only a DC power bus and communication interface from the vehicle.

## 1. INTRODUCTION

Electrification of the driveline of agricultural vehicles, and self-propelled field machinery in particular, enables new flexibility in machine design, along with often lower energy losses compared to hydrostatic drive systems [1]. Classic driveline configurations rely on a cost-effectively mix of mechanical and hydrostatic components: while infinitely variable speed control for both process and ground drive is typically achieved hydraulically. This enables high-level process optimization, but at the cost of considerable power losses, particularly at partial load. Hydrostatic circuits with multiple motors, long pipe runs, hoses, connecting elements and control valves all contribute to these losses. Electric power transmission elements offer substantially lower losses by comparison, and electric motors have better partial-load efficiency than equivalent hydraulic motors.

Regarding basic system functions, electric variable drives are not too far away from to direct hydrostatic drives (without power split). The first known design for tractors was introduced by the Silsoe Research Institute in 1954 [1]. A variable displacement pump was directly driven by the engine crankshaft and large hydraulic radial piston motors were mounted directly on the rear wheel rims. This complete removal of gearwheels, shafts and clutches gave maximal freedom for the frame design, to route pipes and hoses as intended. However, series production did not take place, due to the very high cost of the large hydrostatic radial motors, the need for a completely new vehicle design, and a lower efficiency compared to conventional gearwheel transmissions (at that time still simple at relative low cost).

This paper focuses on two design groups of direct electric drivelines without power split: *axle modules* and *wheel modules*, which are of particular relevance for self-propelled agricultural machinery and equipment carriers.

The standard European tractor, in Europe characterized by four-wheel drive, large rear wheels, smaller steerable front wheels, a cabin mounted near the rear axle and an engine in the front, has been the dominant design paradigm since the mid-20th century, initially with dominating rear-wheel drive. For this vehicle type, the high cumulative cost of distributed electric motors makes axle and wheel module concepts less attractive than central motor packages: a standard tractor with distributed drives may require up to six electric motors (two or four for ground drive, at least one for the rear PTO and one for hydraulics), each with its own power electronics and reductions, resulting in total invested motor power of up to approximately 400% of nominal vehicle power, compared to 100–230 % for central motor configurations [2]. This paper therefore focuses on vehicle architectures where modular driveline concepts offer the greatest benefit: self-propelled agricultural machinery, equipment carrier, propelled trailers and field robots. A companion paper addressing electric driveline concepts specifically for standard agricultural tractors is in preparation.

The range of non-standard vehicles for which modular driveline concepts are relevant is broad. In self-propelled forage harvesters (such as the Krone BiG X, Claas Jaguar, John Deere 9000, New Holland FR and Fendt Katana series), the chopping process consumes the majority of engine power while ground drive requires only a fraction, making modular electric ground drives particularly attractive. Self-propelled combine harvesters (Claas Lexion, John Deere X9, New Holland CR, Fendt IDEAL) present a similar separation of process and ground drive power. Self-propelled sugar beet harvesters such as the Holmer Terra Dos, Ropa Tiger and Grimme Rexor, and potato harvesters such as the Grimme Ventor and Varitron, Ropa Keiler, Dewulf Kwatro and AVR Puma, operate often in wet soil where individual torque control of the wheels supports compaction management. Equipment carriers such as the Nexat [3], Claas Xerion [4] and Holmer Terra Variant, the Krone BiG M self-propelled mower, and self-propelled slurry applicators such as the Vervaet Hydro Trike further illustrate the diversity of non-standard vehicle architectures. Notably, several of these machines already employ hydrostatic individual wheel or axle drives: the Holmer Terra Variant 435 and Terra Dos use individual axle drives without cardan shafts, the Krone BiG M uses individual wheel motors, and the Nexat employs diesel-electric wheel motors, demonstrating that the transition from centralized mechanical drivelines to distributed drives is already underway in this sector.

## 2. REQUIREMENTS

### 2.1. Power demand and load spectra

As noted in the Introduction, the split in power demand between process drive and ground drive is a fundamental design principle for self-propelled agricultural machinery. In forage harvesters and combine harvesters, the process drive (chopping, threshing, and separation) typically consumes 60–80% of total engine power, while the ground drive requires only 20–40%. Both figures apply to typical working conditions in relatively flat terrain; in root crop harvesters, the process power share is somewhat lower but still substantial. This separation of process and ground drive power is a key reason why modular electric ground drives are considered for these vehicles: the ground drive can be electrified independently of the process drive.

A constant power dimensioning is insufficient for dimensioning an electric ground drive, as special conditions impose substantially higher short-duration peak demands. Comprehensive load spectra over the full service life of self-propelled agricultural machinery are rarely published; however, hydraulic pressure measurements taken on a Claas combine harvester over 1,124 hours of evaluation with alternative hydraulic oils provide a certain information, **Figure 1** [6]. The statistic pressure distribution can be interpreted as a superposition of two distinct statistical processes: for the majority of operating time, pressures are between about 50 and 200 bar, representing typical field conditions (broken curve). In addition, there are pressures up to approximately 450 bar, however, 450 bar for only less than 0.1% of total operating time, corresponding to events such as accelerating a heavily loaded vehicle uphill on hard surface. Although wheel torques are not always directly proportional to hydraulic pressures in the case of range transmissions downstream, the shape of this combined load spectrum is instructive for the dimensioning of alternative drive systems.

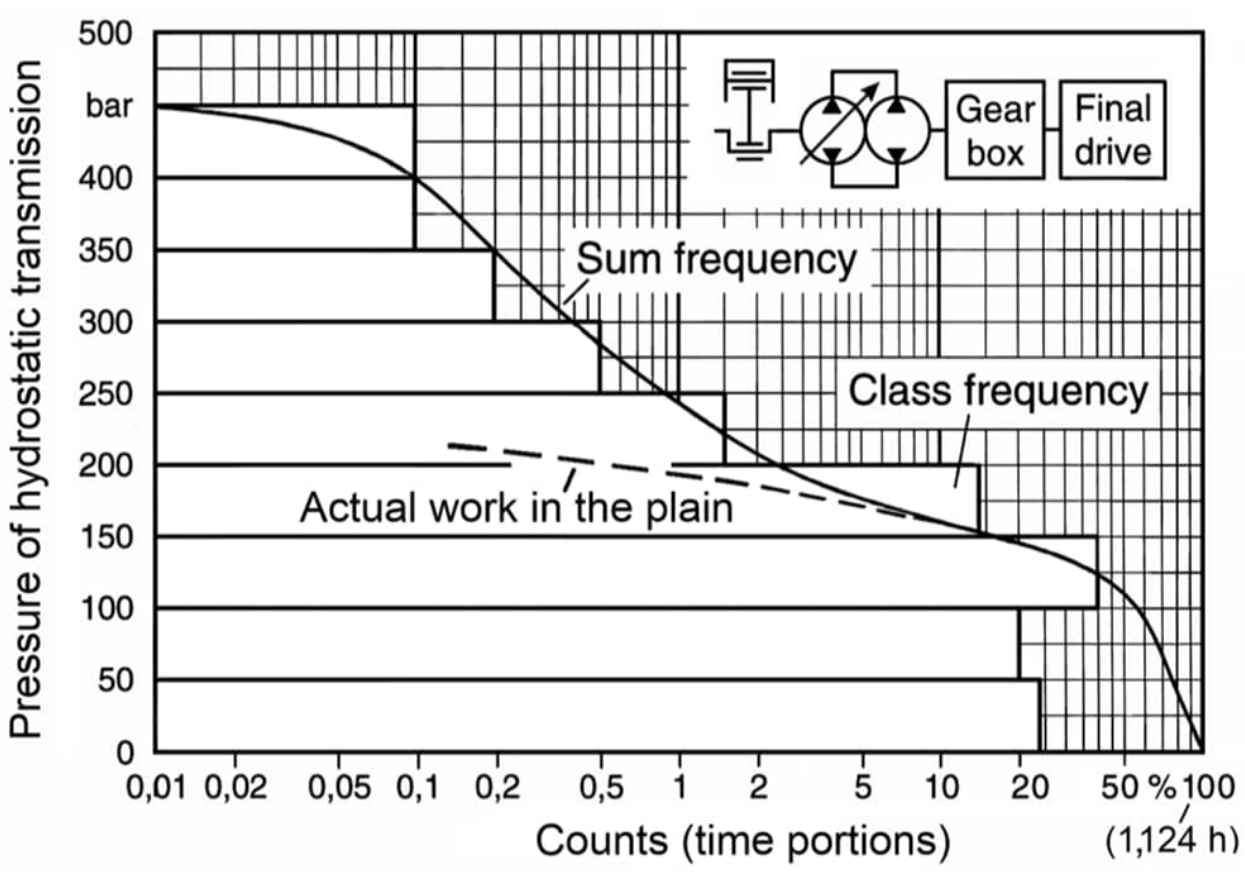

***Figure 1.*** *Hydraulic pressure distribution measured on a Claas combine harvester over 1,124 hours of operation to test alternative hydraulic oils [6]. The combined distribution reflects a superposition of typical working conditions (50–200 bar, broken curve) and high-pressure events up to about 450 bar, however, 450 bar during less than 0.1% of total operating time*

Measurements at the Krone BiG X 1000 forage harvester confirmed that the chopping process consumed over 70% of installed engine power while the hydrostatic traction drive required only approximately 10% during field operation at 8 km/h [5]. Higher torque loads were recorded under special operating conditions, and reverse-mode loads were significant at approximately 13% of total operating time. Short-duration overload events of this character can be expected as typical for heavy self-propelled machinery in general, requiring electric motors capable of delivering high short-time peak power.

### 2.2. Efficiency

With the power of self-propelled agricultural machinery – today typically rating between 200 and 750 kW, the efficiency of the ground drive is more important than it was for earlier, lower-powered machines. The efficiency advantage of electric over hydrostatic axle drives has been quantified experimentally. Heckmann [7] built a diesel-electric rear axle prototype for the Krone BiG X forage harvester, replacing the hydrostatic drive (by radial piston motors) with electric machines using permanent-magnet synchronous motors and planetary gears at each wheel hub. Under typical field conditions (5 km/h, 4.5 kNm wheel torque), the electric axle achieved an efficiency advantage of 17 percentage points over the hydrostatic series drive. When evaluated over a complete field-derived load cycle, the advantage increased to 23–27 percentage points, because the hydrostatic system suffers disproportionately at the partial loads that dominate field operation.

In spite of this, efficiency under full load remains important. **Figure 2** presents a well-known efficiency target for tractor CVT transmissions under full load, developed for power-split CVTs of large tractors around 1991/92 for Fendt and CLAAS and later released for publication [1]. Present direct hydrostatic drives are by far not able to meet this target. The working speed range is for most self-propelled harvesters typically 3–8 km/h for field operations, which is narrower than the 6–12 km/h focus area shown in Figure 2 for standard tractors.

The same principle has been applied by Bernhard in 2011 to evaluate the ground drive of self-propelled agricultural machinery [8]. He proposed an efficiency target function specifically for the combine harvester ground drive, derived from measured operating profiles, see **Figure 3**. This target differs considerably from the tractor CVT target regarding two criteria: ten points lower in efficiency and with a slightly lower main working speed range due to the special working conditions. Optimized electric drives of axle or wheel modules can potentially exceed the Bernhard target considerably, perhaps even approaching the Renius efficiency target.

*2.3. Speed spectra and spread*

Self-propelled agricultural machinery operates in a narrower working speed range than standard tractors. Field work speeds are typically 3–8 km/h for combine and forage harvesters, and 4–7 km/h for root crop harvesters. Road transport speeds range up to 25 to 40 km/h depending on vehicle type and regional market requirements. The overall speed ratio from minimum field speed to maximum road speed is therefore approximately 5:1 to 13:1, substantially less than the 40:1, or more, required for standard tractors (maximum often 32 or 40 km/h).

Equipment carriers present a wider speed range than self-propelled machinery, as they may be used for tillage, seeding and transport operations on road. Range requirements are therefore closer to those of standard tractors.

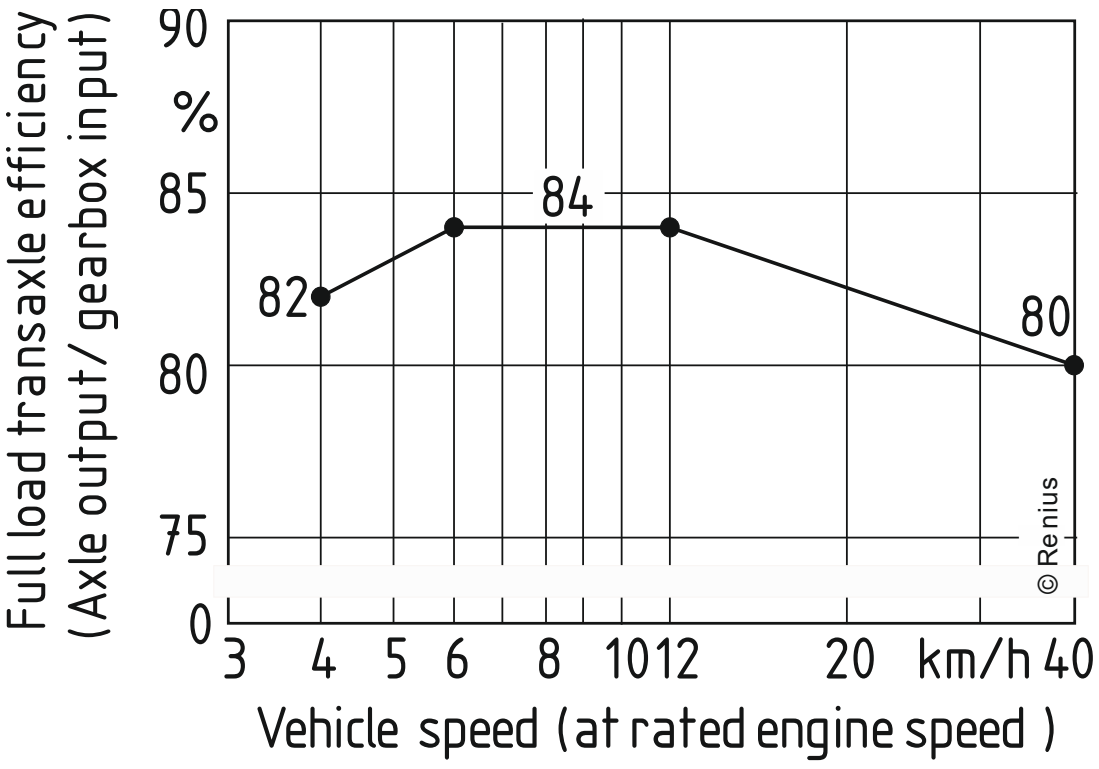

***Figure 2.*** *Full load efficiency target (axle output/gearbox input) for tractor CVTs of nominal engine power above about 100 kW, details see Renius [1]*

For the driveline design, the narrower speed range has significant implications for the number of ranges required. A self-propelled machine with a maximum road speed of 25 km/h, such as the Holmer Terra Dos sugar beet harvester, requires a speed ratio of approximately 5:1, which may be achievable with a single range. At 40 km/h the speed ratio increases to approximately 13:1, typically requiring two ranges. Field robots which are not designed for self-propelling on public roads have a speed spread approximatively 8.

*2.4. Manufacturing*

Self-propelled agricultural machinery is produced in substantially smaller series than standard tractors. This has important implications for the driveline design.

For small-series, the development and tooling cost per unit is high, and the use of standardized or modular driveline components from specialized suppliers is economically attractive. The axle module and wheel module concepts are particularly well suited to this manufacturing context: a specialist supplier can develop and produce the module at volume with economies of scale, while the vehicle manufacturer usually cannot. They can focus on the process-specific part of the machine and the frame design. This division of engineering responsibility reduces the total investment cost and allows smaller manufacturers to start with high-quality electric driveline technology under the same overall strategy.

Manufacturing costs include component costs, custom-made elements such as housings, and assembly expenses. Reduction of the number of unique components simplifies production and servicing. In this context, identical axle or wheel modules used across multiple vehicle platforms, or even across multiple manufacturers, can provide significant economies of scale.

*2.5. Other requirements*

In addition to efficiency, speed range and manufacturing, several further requirements must be considered for driveline design of self-propelled agricultural machinery,

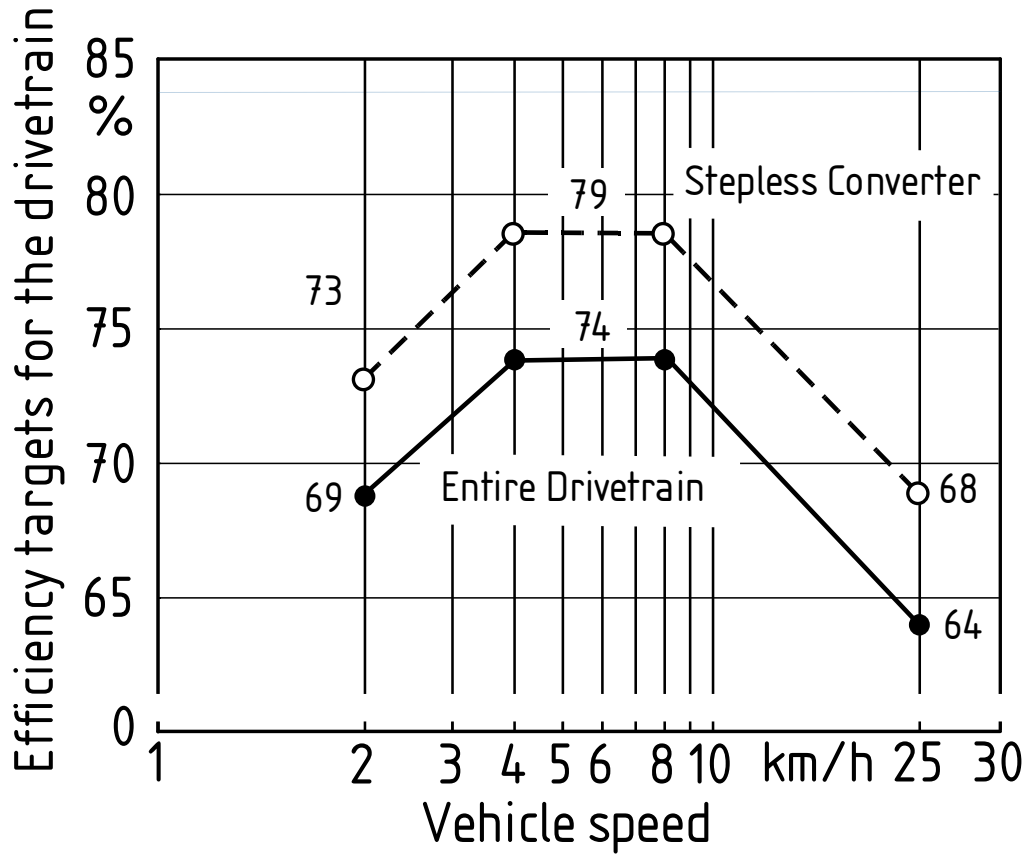

***Figure 3.*** *Efficiency target function for the combine harvester ground drive during harvest. Adapted from Bernhard [8]*

including safety, soil compaction, reliability and serviceability and – of course – first cost.

Unlike standard tractors, where all driven functions share a single integrated driveline optimized for one primary use case, self-propelled agricultural machinery benefits from separating the process and ground drives. In axle and wheel module architectures, the modules handle ground drive exclusively, while process drives are powered independently. This separation allows each drive to be individually optimized and enables, for example, the ground drive to be electrified without requiring changes to the process drive architecture.

Safety and regulatory compliance impose requirements on driveline design. Road homologation requires service braking and parking braking. Their design must meet regional standards. The braking system must function independently of the electric drive system to ensure fail-safe operation. Functional safety standards require that loss of electric power does not result in uncontrolled vehicle movement or loss of steering function.

Soil compaction is a critical concern for self-propelled agricultural machinery, which is typically much heavier than standard tractors. Self-propelled sugar beet harvesters can exceed 60 tons when loaded. Minimizing soil compaction is therefore a significant requirement, addressed through the use of very large tires, tire inflation pressure management, optimal weight distribution, controlled traffic patterns and precise torque control per wheel. Modular electric drives with individual wheel control offer potential advantages in this regard. Rubber tracks can also reduce compaction.

Reliability and serviceability are critical in agricultural applications, where downtime during peak season has significant economic consequences. Self-propelled harvesters typically operate intensively during short seasonal windows (e.g. for only 2–4 weeks at sugar beet harvest), and any driveline failure during this period can result in substantial crop and income losses.

### *2.6. Sources of electric energy*

Both module types require only a single power input, a DC bus, combined with integrated power electronics to drive the AC motors at variable frequency for infinitely variable speed. The primary power source may be a battery, a combustion engine driven generator or a fuel cell.

For large self-propelled machines, an on-board battery of sufficient capacity would be extremely large, heavy and expensive [2]. This is one reason why the Nexat Gantry [3] and large field robots use combustion engine driven generators as the primary power source. Small field robots and lighter autonomous machines are well suited to battery-electric operation, and some incorporate solar panels as a supplementary energy source [2].

## 3. ELECTRIC DRIVELINE CONCEPTS

Two design groups of direct electric drivelines without power split are considered in this paper: axle modules and wheel modules.

In the axle module concept, each axle is a self-contained mechatronic unit comprising the electric motor/machine, transmission, power electronics, control electronics, integrated steering and braking system. A vehicle designer assembles two or more such modules with a frame to create the complete vehicle.

In the wheel module concept, each wheel has its own independent module containing the full electric machines, transmission, power electronics, control electronics, integrated steering and braking system. This provides maximum freedom for frame design, as the space between the wheels can be used entirely for process equipment, which is a decisive advantage for self-propelled harvesters, equipment carriers and other machines where the vehicle carries integrated process functions.

In addition to drive for propulsion, both the axle module and wheel module concepts include integrated steering actuation. Integration of steering actuators into front axles was started in the early 1980s (ZF and SIGE-Deutz) [1], but these remain partial modules, as the hydraulic steering valve is not an integral part of the axle. The modules proposed herein feature integrated steering mechatronics, which allows the whole module (axle or wheel) to have only one power input (a DC bus) with integrated power electronics. The concepts do not define how steering actuation is implemented, as there are several mechatronic options, including the use of hydraulics as part of the system.

## 4. WHEEL MODULES

The first known attempt to realize hydrostatic wheel hub drive on a tractor of conventional layout was made by the Silsoe Research Institute, UK, in 1954, **Figure 4**. This prototype was based on two radial piston motors in the rear wheels while hydrostatic power was produced with a variable displacement pump driven by the engine crankshaft.

While radial piston motors were used to realize wheel hub motors, this does not meet the definition of the wheel module used herein, as the control was in the pump, not in the motor (control of the motor would have required a constant pressure system). Hydrostatic drives have been successfully used in several other off-road vehicles, such as excavators and other earth moving machinery.

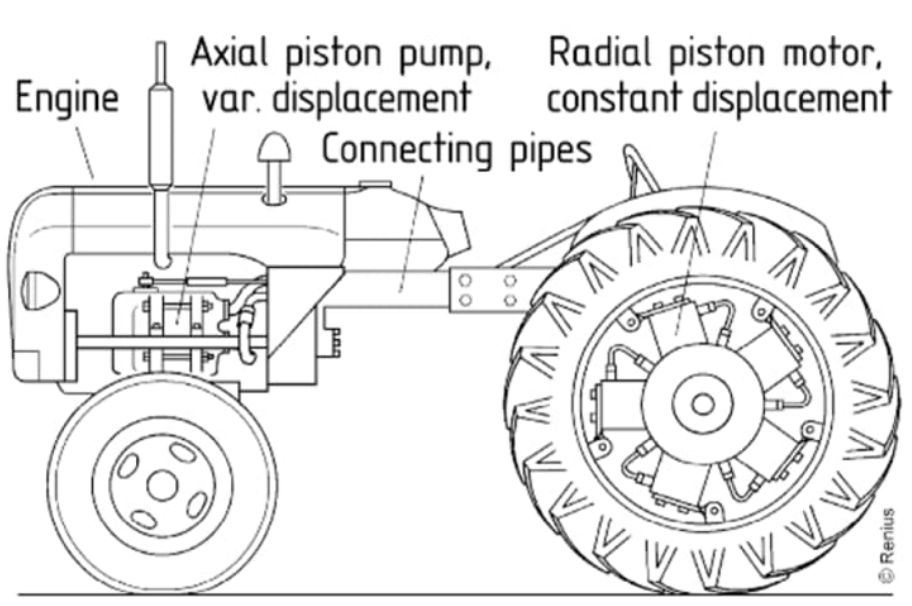

***Figure 4.*** *Direct variable hydrostatic ground drive by NIAE at Silsoe in 1954 [1]*

Similarly as in the Silsoe Research Tractor, wheel hub motors were also used in the unmanned tractor prototype made by Modulaire Ltd., in Finland, developed in 1990-1995, **Figure 5**. This tractor design has four hydrostatic radial piston motors, one for each wheel; and four independently electronically controlled variable displacement pumps, one for each wheel. [9]

This four-wheel driven and four-wheel steerable vehicle had in total eight degrees of freedom for motion, as each wheel was driven and steered independently, enabling high precision Ackermann drive with each wheel at theoretically correct angle and speed around the intended center of rotation, offering track-in-track drive and pull-in-turn capability. Dynamic control of the whole system was difficult. Also this design does not meet the requirements of the wheel module definition used herein, as control is again in the pumps centralized, not in the wheels itself. [9], [10], [12]

With electric drives, the system makes more sense. A typical layout of electrified drives consists of a DC power bus, with super capacitors or batteries as buffer and each actuator being able to operate within control limits. This DC power bus approach enables the idea of the wheel module, so that not only the electric motor is part of the wheel hub, but also the power electronics.

A proposal by Renius [1] to realize a wheel module for agricultural tractor is presented in **Figure 6**, based on the idea to include two ranges into the wheel combined with a high speed reduction, realized with a special planetary gear arrangement. The first planetary reduction serves only to enable high motor speeds, reducing motor dimensions and cost. The second stage provides two ranges shifted through externally controllable brakes. Their position avoids the need for hydraulic lines carrying control and cooling pressure into the rotating system. In the second stage, the planetary carrier is connected to the wheel hub and the ring gear is fixed, the typical configuration in tractor final drives. This arrangement offers overall speed reductions of up to about 60:1 in low range and 150:1 in high range.

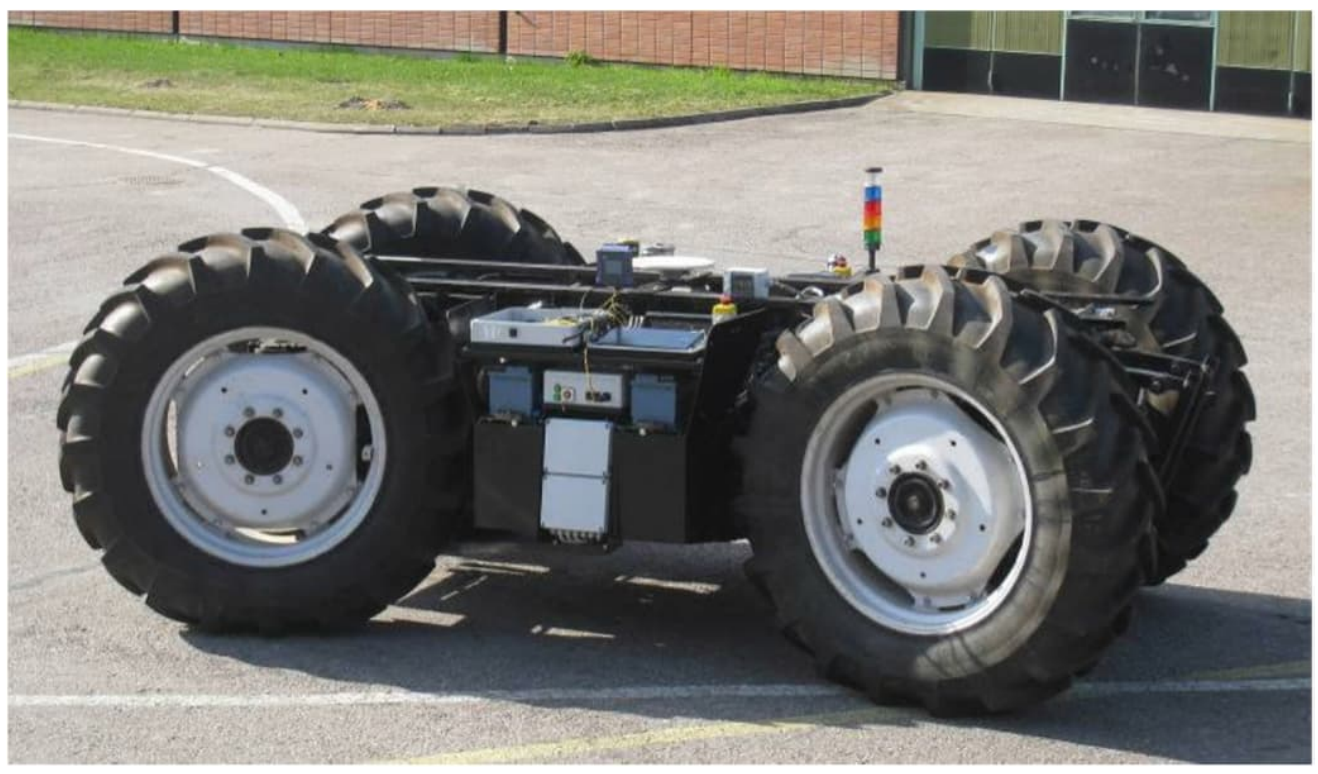

***Figure 5.*** *The Module, developed for unmanned operations by Modulaire in 1990; later refurbished and automated by Oksanen [11],[12]. This tractor prototype has fixed displacement radial piston motors in each wheel hubs and driven by four variable displacement pumps, each separately controllable. Radial piston motors are the hydrostatic realisation of wheel hub motors [9]*

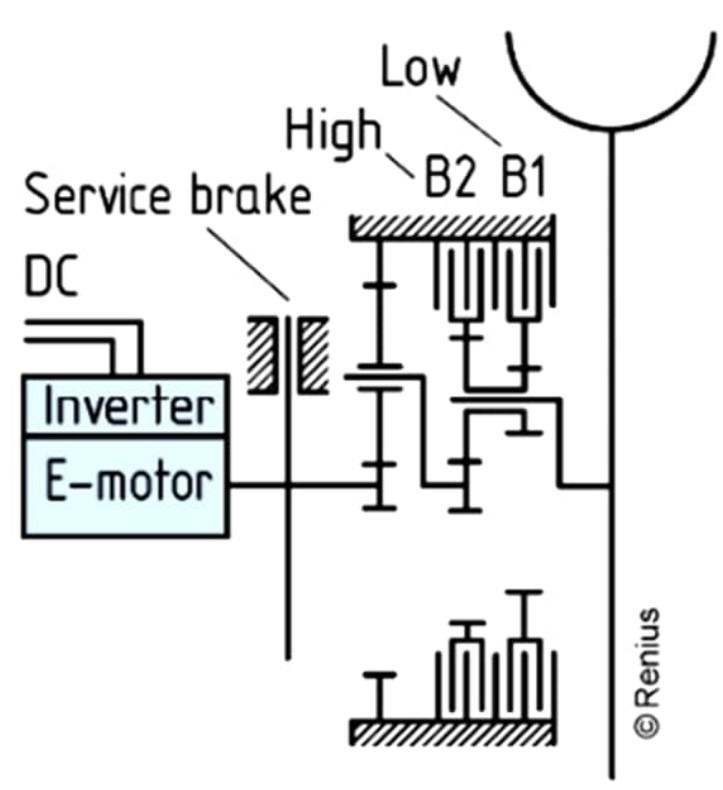

***Figure 6.*** *Electric wheel drive with two power-shifted ranges. Proposal of Karl Th. Renius in 2020 [1]*

A service brake is between electric motor and reduction, where speed is high and torque low. The brake can also be part of the electric motor assembly. This brake is required for regional road regulations. An additional parking brake can be realized at almost no extra cost by closing both brakes B1 and B2, fixing both planetary ring gears and blocking any rotation of the planetary carrier.

Proper engineering of the parking brake feature requires designing this element so that regional regulations are fulfilled. Alternatively, the parking brake can be realized in location of the service brake, but at higher cost.

## 5. AXLE MODULES

The concept of axle modules discussed in this paper means that an integrated system contains electric machine, transmission, brakes, power electronics, control electronics and optionally steering are in one package. A specialist supplier manufactures the module; a vehicle designer assembles two or more such modules with a frame to realize the complete vehicle.

### *5.1. Axle modules of small field robots*

The idea of the axle module was born on 5 August 2008, when the lessons learned from the 2008 field robot event were evaluated. The concept was subsequently tested in several field robot designs that attended the annual Field Robot Event: 2009 (EasyWheels), 2010 (TurtleBeetle), 2011 (Cornivore), 2012 (RoseRunner), 2013 (DoubleTrouble) and 2015 (GroundBreaker).

The primary motivation was to encapsulate the drive design independently of the frame design, enabling parallel development and modular manufacturing. An additional reason of identically sized units was to allow rapid substitution of a unit in a case of break down (three were manufactured, one for a spare part). **Figure 7** shows the structure around 2009.

This interchangeability was recognized at the Field Robot Event 2009 as a useful key feature of the concept.

The robot mass was between 10 and 20 kg, and components were scaled accordingly. Each axle module integrated brushed 12 V DC motors for propulsion and a high torque servo for steering, with all power electronics housed within the aluminum enclosure; the H-bridge MOSFET devices were mounted directly on the enclosure walls, using the aluminum housing as a passive heat sink. Axle modules communicated with a central controller via CAN bus. Speed control, torque control and steering position control were all implemented within the module itself. A typical appearance of an axle module is shown in Figure 7 (it is the first generation model from 2009, EasyWheels).

**Figure 8** shows a possible internal drive arrangement using widely available components. Replacement of the bevel gearing with spur gears improves efficiency, compensating for the additional planetary stage.

### *5.2. Modular axles for on-road vehicles*

Although not directly relevant for agricultural machinery, electric e-axle modules for on-road trucks have a structure that is architecturally similar to possible solutions for agricultural field machinery. ZF has, for example, developed the concept under the name AxTrax2, released in 2023 and in production since 2024 [14]. It is intended for trucks up to 44 tons, **Figure 9**. Even though it does not contain steering, it is structurally similar to the design realized for the field robot in 2008–2009 (Figure 7). Final planetary reductions in the wheel hubs are, for example, very usual in tractor and construction machinery front axles.

The concept of integrated electric axle drives has gained significant momentum in the commercial vehicle sector. In addition to ZF AxTrax2, several other manufacturers have developed comparable e-axle systems for trucks. Dana has introduced the Spicer Electrified eS9000r e-axle for Class 4 and 5 commercial vehicles, integrating a TM4 SUMO LD electric motor and inverter into a single axle unit with a maximum power of 237 kW [15]. Allison Transmission launched their eGen Power family of fully integrated electric axles, with the heavy-duty eGen Power 100D variant providing over 450 kW of continuous power from two integrated electric motors for Class 8 trucks [16]. BorgWarner has developed the integrated drive module (iDM), combining electric motor, power electronics and transmission in one e-axle unit, available for light commercial vehicles up to 7.5 tons [17]. These commercial vehicle e-axle products represent significant steps towards the axle module concept.

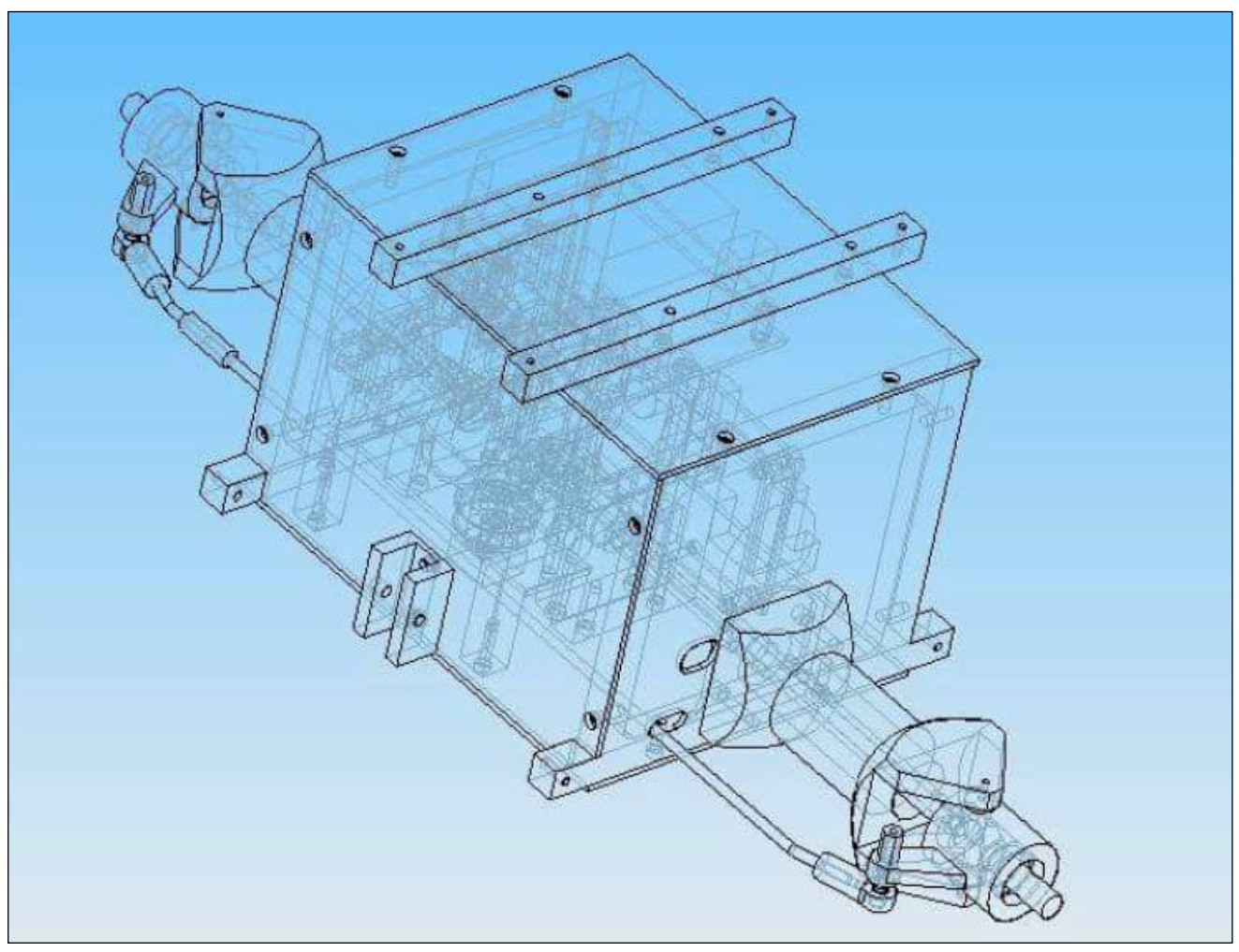

***Figure 7.*** *Axle module 2009 [13]*

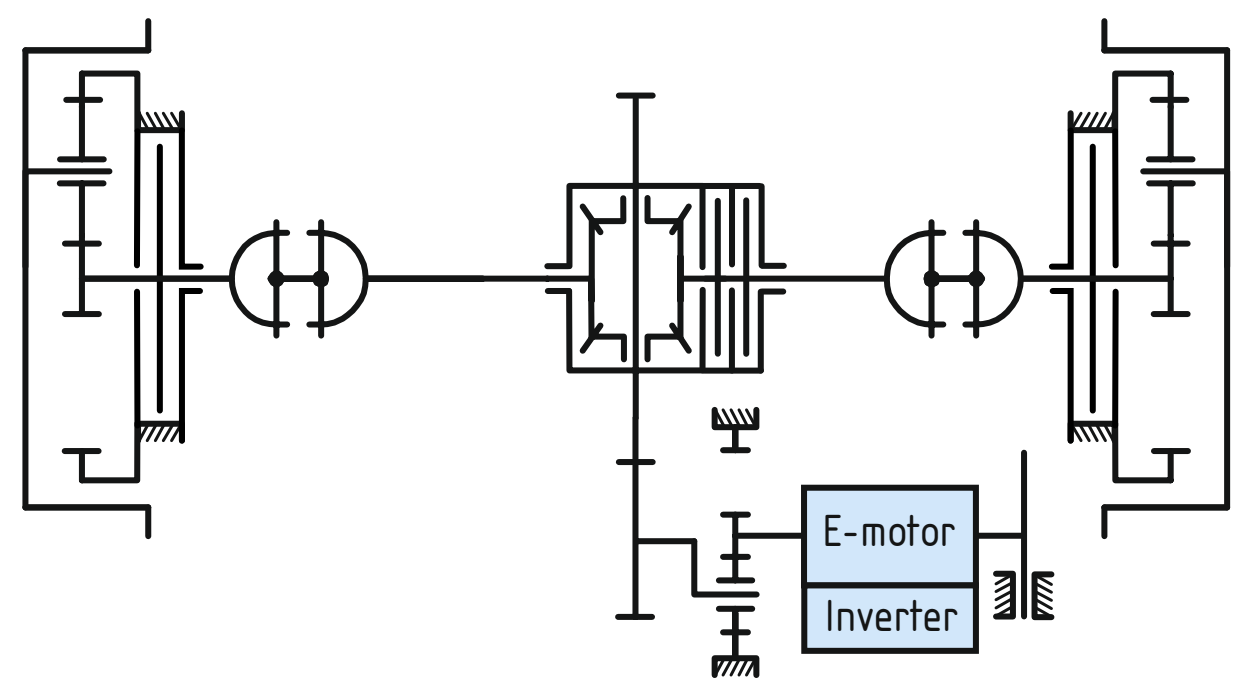

***Figure 8.*** *Axle module, one possible design using existing elements. Integrated steering elements not shown*

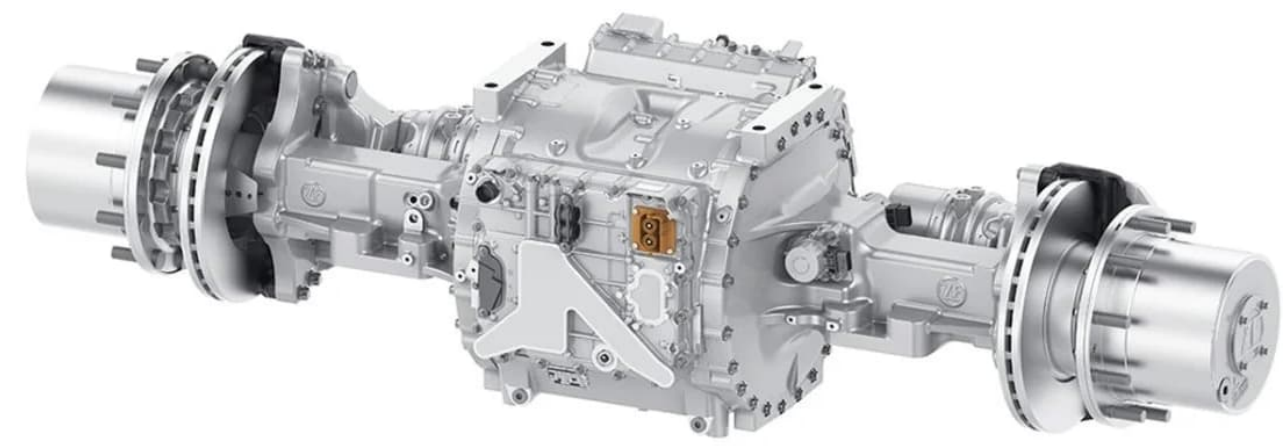

***Figure 9.*** *ZF AxTrax2 for trucks (2023) [14]*

However, none of them integrates steering into the axle unit, and the control intelligence typically resides in a central vehicle ECU rather than being distributed into the axle itself. In the definition used in this paper, the axle module is a truly self-contained mechatronic unit with integrated power electronics, control electronics and steering, so that a vehicle manufacturer can build a complete vehicle by assembling two or three such modules with a frame, requiring only a DC power bus and a high-level communication interface. The truck e-axle products mentioned above are architecturally similar, but do not yet achieve this level of modularity and distributed intelligence completely.

### *5.3. Off-road electric drives*

In the off-road and mining sector, diesel-electric wheel drives have been standard practice for ultra-class haul trucks for several decades. Komatsu's 930E series, first introduced in 1996, was the first haul truck to employ AC electric traction motors, with GDY106C wheel motors providing propulsion for a 290-tonne payload capacity vehicle [18]. Caterpillar's 795F AC mining truck similarly employs AC induction wheel motors at 2,600 V in a high-voltage, low-current configuration [19]. Liebherr's T 264 uses the Litronic Plus AC drive system with forced-air-cooled AC induction wheel motors and 40:1 planetary final drive reduction, designed for 240-tonne payloads [20]. These mining truck drives are the most mature large-scale application of electric motors at the wheel. Architecturally, these are wheel drives rather than axle drives: each rear wheel has its own electric motor and planetary final reduction, without a solid axle housing connecting left and right sides. However, they do not constitute wheel modules in the sense of this paper: the control system is centralized, the power electronics are located remotely from the motors, and steering is not integrated. The distinction is important: electric wheel motors with a central ECU have existed for decades, but the wheel module concept proposed herein requires that each module carries its own power electronics and control intelligence, making it a self-contained unit.

### 5.4. Agricultural electrification initiatives

The agricultural sector has also seen several electrification initiatives relevant to this discussion. The Nexat [3], awarded by a Gold Medal at Agritechnica 2022, represents an advanced equipment carrier concept with four electrically driven large wheels and interchangeable implements for tillage, seeding and harvesting. Monarch Tractor has developed a fully electric autonomous tractor for vineyard and orchard applications, and Solectrac offers battery-electric tractors for smaller agricultural operations. John Deere demonstrated the GridCON concept, a cable-connected fully electric tractor.

These developments indicate growing industry interest in electric driveline concepts for agriculture.

In a research project, Heckmann [21] implemented and tested a diesel-electric rear axle drive on a Krone BiG X forage harvester, replacing the series hydrostatic wheel motors with permanent-magnet synchronous motors and planetary gears; this represents the most comprehensive published comparison of hydrostatic and electric axle drives on a self-propelled agricultural machine.

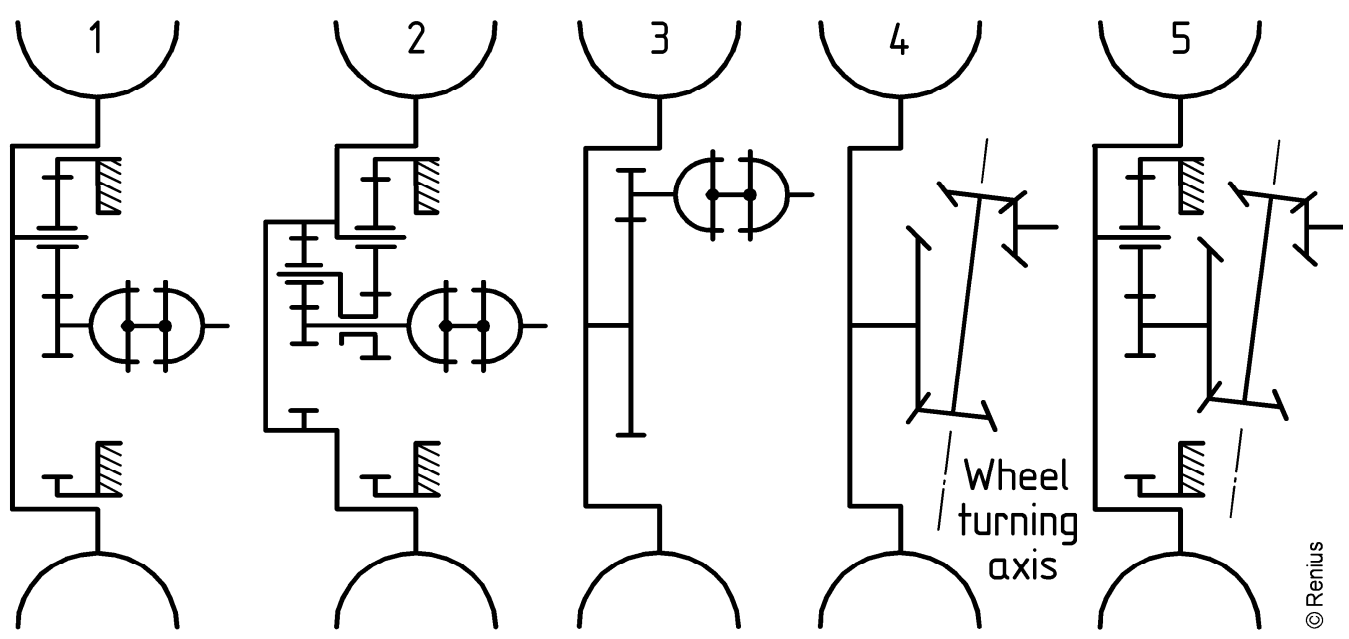

***Figure 10.*** *Basic final drives of driven front axles: only one side plotted and differential not shown. Design 1 is the most important*

However, none of the current production or prototype agricultural vehicles implements the axle module or wheel module concept with truly distributed control as defined in this paper.

## 6. COMPARISON OF MODULE TYPES

### 6.1. Space and ground clearance

Wheel modules do not restrict the space between left and right wheels and therefore allow high freedom of design. Axle modules provide structure to support the vertical wheel loads, either through a kingpin or through a solid axle housing when steering is not integrated.

Examples of such high loads are front axles of combine harvesters. The cutting platform imposes significant vertical and longitudinal loads on the front of the vehicle. In forage harvesters and sugar beet harvesters, the harvesting head or lifting unit is similarly front-mounted and heavy. Additional loads must be carried regarding the heavy process equipment. The axle module concept is well suited for carrying such loads. However, axle modules reduce the design freedom between the wheels considerably. Wheel modules offer this freedom, the loads must be carried entirely by the frame or suspension structure between the two independent wheel modules, which requires additional structural elements and carefully engineering of the load paths.

Wheel modules additionally allow complete freedom in the longitudinal center of the vehicle between the axle positions. For self-propelled harvesters, large crop flow channels, threshing elements and elevators must pass through the central section of the machine, the absence of a transverse axle housing is an important design advantage that can simplify the internal layout and improve ground clearance in the central section.

Axle modules limit ground clearance similarly as driven axles in current agricultural vehicles. The axle module can be realized with all basic final drives used today in driven front axles, see **Figure 10**. Only one side is plotted and the differential is not shown. For smaller vehicles, final drive portal axle designs 3–5 are often preferred to provide sufficiently high ground clearance, while for larger vehicles designs 1–2 are preferred with a clear priority of design 1.

### 6.2. Suspension

In conventional configurations, a solid axle module has significant unsprung mass, together with wheels and tires. To reduce the unsprung mass, the housing of the axle module can be designed much smaller if it is connected with the wheels by a double wishbone, **Figure 11**. This design principle is already established in tractor front axles, e.g. by tractors of Case and Steyr since 1999 [1]. It allows a further use of the king pin connection with the chassis (if needed) when introduced with suspension components [1].

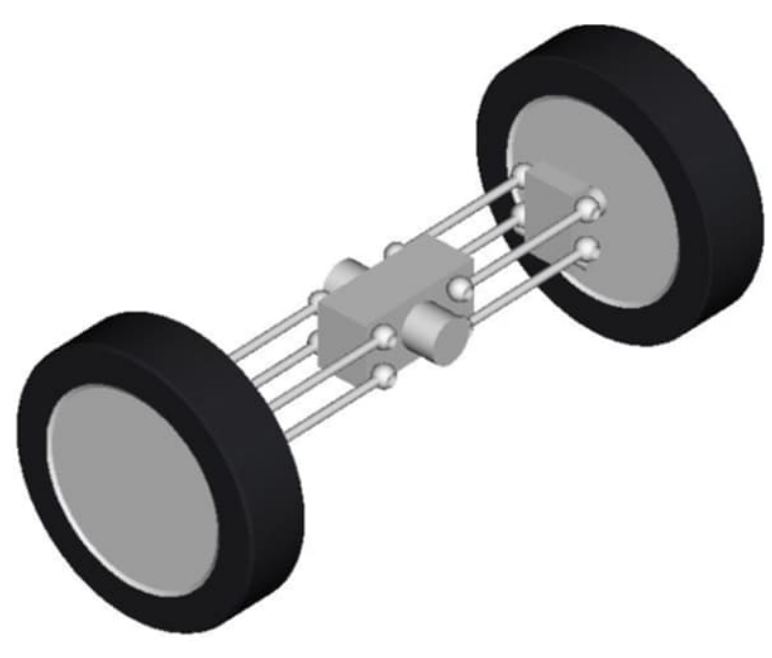

***Figure 11.*** *Single wheel suspension by double wishbone linkage, combined with a king pin connection [1]*

This design allows realization of axle module with suspension components, however, with some complications regarding wheel connection and the steering.

Wheel modules allow maximum design freedom for suspension as left and right wheel modules are not coupled. Wheel modules do not only allow independent suspension by wishbones, but also alternative frame designs such as seen in Nexat vehicles [3], or in self-propelled sprayers. However, a wheel module has a fixed unsprung mass, with an electric motor and transmission, a wheel and a tire, and this is a likely higher weight than with double wishbone axle module.

Both wheel modules and axle modules need to provide fixture points for the frame or suspension systems of the vehicle.

An example for a fully suspended field robot is shown in **Figure 12**, where load balance and suspension are realized with a multi-link mechanism and spring-damper elements. Axle module provides three ball joint connections to support three degrees of freedom (one at bottom, two on top, forming a triangle), and two additional ones for vertical load (bottom corners) [13].

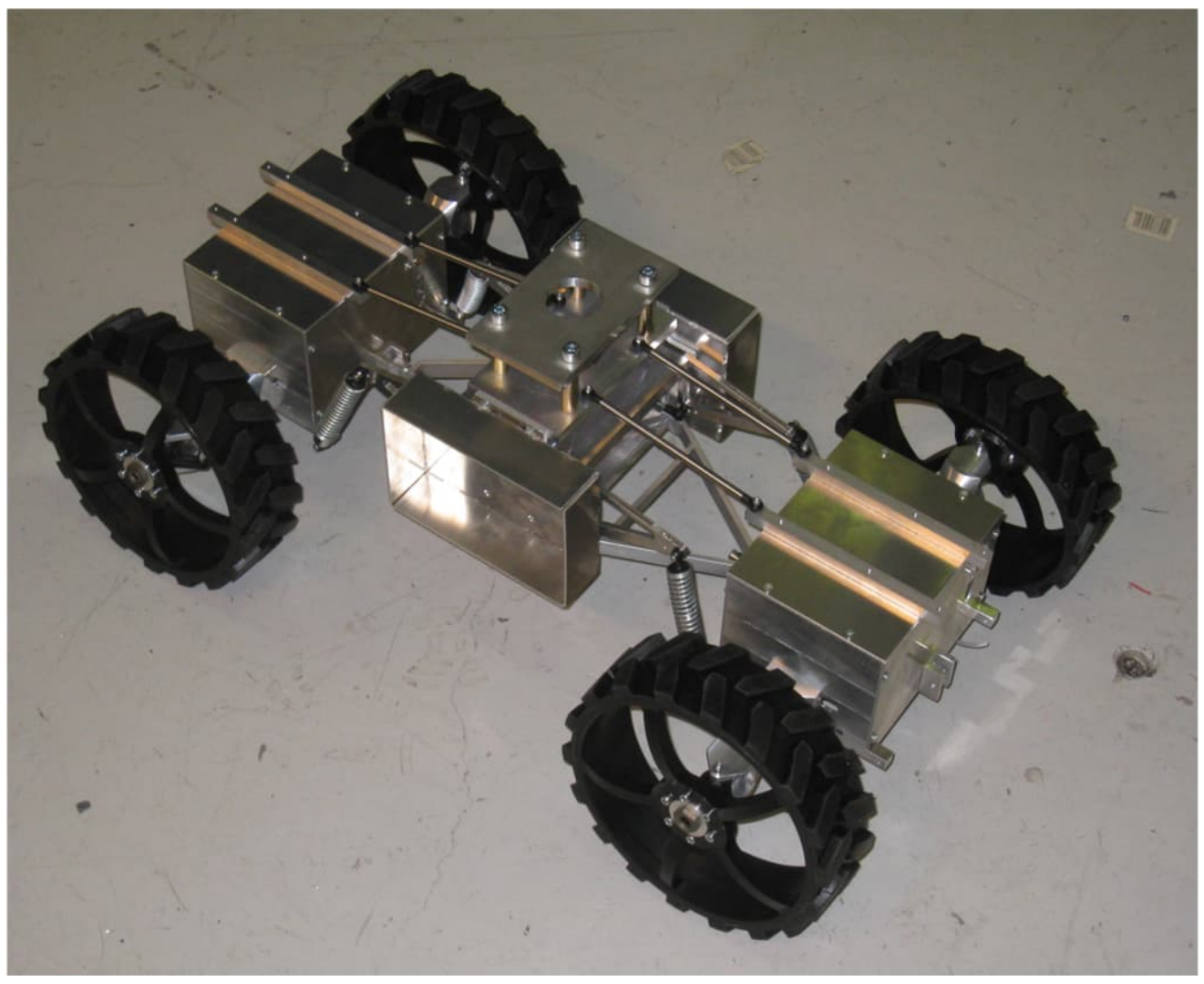

***Figure 12.*** *Frame design based on two axle modules [13]*

This interface design illustrates that axle module needs several fixture points, so there is one significant difference to conventional front axle design of tractors.

### *6.3. Manufacturing and development costs*

Based on engineering experience, the manufacturing cost of one solid axle module, with a single electric motor, is estimated to be notably more economical than two wheel modules providing the same performance. Quantitative data on the cost of electric versus hydrostatic axle drives is scarce. Heckmann [7] reported that the electric rear axle prototype for a forage harvester cost approximately 4.5 times more than the hydrostatic series axle (in 2016: 29.5 k€ versus 6.5 k€ at an assumed production volume of 100 units per year), and the electric drive doubled the axle weight (weight-to-power ratio changing from 18.2 to 36.3 kg/kW). Amortization through fuel savings alone was not achievable within the useful life of the machine. The study concluded that cost parity requires wide adoption of electric drives across the commercial vehicle industry, enabling cost-effective modular components, which aligns with the cross-platform standardization argument for axle and wheel modules presented in this paper.

Development costs highly depend on the starting point. For instance, for a manufacturer with engineering experience in driven front axles or hydrostatic axle drives, selecting appropriate transmission components for the axle module could be developed based on earlier designs and test data.

An important consideration for electric motor dimensioning is asymmetric vertical wheel load. Self-propelled agricultural machinery frequently operates on sloping terrain; sugar beet and potato harvesters working on side-hills, for instance, or combine harvesters traversing undulating fields. In such conditions, the downhill wheels bear a disproportionate share of the vehicle weight, which is particularly significant for heavy machines exceeding 30 tons. Higher vertical load results in higher available and required traction force, demanding more torque from the drive. In the axle module concept, the differential distributes torque automatically between left and right wheels, and the single electric motor only needs to provide the total axle torque. In the wheel module concept, however, each motor must be dimensioned individually for the worst-case single-wheel load condition. This means each wheel motor must be rated for significantly more than half the equivalent axle motor torque, leading to overdimensioning for normal symmetric operation. This asymmetric load effect is a notable disadvantage of wheel modules in terms of motor cost, weight and packaging, and is one of the reasons the electric motor dimensioning favors the axle module concept.

However, test bench development for the wheel module is potentially cheaper, as only one power brake unit is required rather than two for the axle module.

*6.4. Steerability*

Axle module concept provides similar approach for steering as tractor front axles since the early 1980's (ZF and SIGE-Deutz): the steering cylinder can be integrated into the axle and design gets more compact [1].

With a single steering cylinder, the left and right wheel angles are coupled through a track rod that can be designed to approximately realize the Ackermann principle. In four-wheel steerable designs, fine adjustment of individual wheel angles may require additional actuators.

Similar to 1980's front axles, the wheel module can have integrated steering capability. This would require that each wheel module incorporates both traction and steering capability in one unit.

*6.5. Controllability*

For the ground drive of two axle agricultural vehicles, axle modules provide two degrees of freedom (plus two additional with steering) and wheel modules provide four degrees of freedom (plus four additional with steering).

While the axle module only provides benefit to drive front and rear axles with different torque or speed (as desired), the wheel modules allow more comprehensive control strategies, like torque vectoring.

With two axle modules, the front and rear axles can be driven at different speeds, enabling pull-in-turn behavior. The turning radius in the field can be noticeably smaller with a pull-in-turn system, which is particularly valuable for headland turning with long self-propelled machines.

Wheel modules provide even greater controllability. With wheel modules, each wheel can be driven separately to enable advanced driving modes. In tight curves, both sides of wheels can be driven with Ackermann or modified Ackermann speeds, providing more accurate traction on both firm and soft soil. On sloping terrain, individual wheel torque control can compensate for asymmetric traction conditions; for instance, reducing slip on the uphill side while maintaining drive on the downhill side. When wheel modules have integrated steering capability, full eight-degree-of-freedom control on sloping terrain can better eliminate drifting downhill while keeping the vehicle heading as intended. For heavy self-propelled harvesters operating on soft soil, torque vectoring also offers the potential to reduce compaction by distributing drive forces more evenly across wheels.

Four-wheel steering is a feature of particular interest for self-propelled agricultural machinery, where maneuverability on the headland is critical for large machines. The CLAAS Xerion equipment carrier has a four-wheel steering since 1997 [4], and several self-propelled harvesters employ multi-axle steering to achieve acceptable turning radii despite their size and weight. With both axle module and wheel module concepts, four-wheel steering can be realized as an integral part of the module design, without requiring additional space in the vehicle chassis.

*6.6. Brake system*

Both axle modules and wheel modules require braking system in addition to the regenerative electric machine. Brakes are required for road regulations, but also parking brake is required to put vehicle on uneven surface unpowered. Preferably both service and parking brake should be realized with the same element.

In both concepts, the brake system shall be wheel-specific. In the axle module, brakes positioned upstream of the differential are generally only permissible in combination with a differential lock.

The proposal of wheel module based on Figure 6 has a clear advantage that the parking brake could be realized by engaging both range brake element together, which prevents rotation of the wheel. This integrated approach of parking brake would save costs. However, realization of this concept requires engineering research, to comply with regulations and reach required braking torque.

In agricultural vehicle design, wet brake elements have been preferred in recent decades for driven axles, positioned between the differential and final planetary drive. The main motivation to use wet brake elements is durability over the lifetime, as this element is difficult to maintain, repair or replace. Some manufacturers have preferred large dry disc brakes near the wheels, while many others use wet disc brakes integrated in the final planetary drive. Dry discs have lower drag torques, important for on-road driving and rather cheaper.

When it comes to axle modules and wheel modules, both dry and wet disc brake options can be used. For instance, ZF AxTrax2 visibly implement dry disc brakes as typical for trucks, but the wheel modules as presented in Figure 6, allows both options depending on the housing.

Changing elements of a dry disc would be easy after uncoupling the electric motor. Wet brakes make sense when located in the same housing with planetary drive, or differential. This is possible also in axle modules.

*6.7. Speed reduction aspects*

In driveline design for self-propelled agricultural machinery, the total speed reduction from electric motor shaft to wheel must be distributed appropriately into components across the driveline. When moving from conventional transmissions to direct electric drivelines (without power-split), this question is still relevant, even if the electric motor has different characteristics.

Any electric motor requires speed reduction in agricultural vehicle applications, where large-diameter wheels must rotate at low speeds compared to feasible electric motor speeds. Extremely high motor speeds, such as 20,000 rpm, are technically feasible and offer a high power-to-weight ratio, but require very high overall reduction ratios which tend to reduce driveline efficiency. A nominal motor speed of approximately 6,000 rpm is therefore considered a reasonable compromise. High torque is required especially at low speeds, starting from standstill under load.

It is a relevant question where speed reduction is done. In the proposed wheel module concept (Figure 6), the advantage is that it is possible with this special two-stage reduction to assemble both the required ranges and the reduction in one package to reduce costs. In the axle module concept, one reduction is in the center of the axle module and another reduction (final reduction) is in the wheel hub, similarly as in typical driven front axles. This is a significant difference.

### *6.8. Free-wheeling*

Multi-axle agricultural vehicles are sometimes designed to disconnect the power train of some axles during road transportation. The main reason is to avoid tire wear on tarmac if wheel circumferential speeds differ between axles. A typical solution uses a clutch, opened for road operation.

In direct electric drive propelled vehicles, this requirement might still be necessary. Decoupling wheel from motor is possible with ranges, like presented in Figure 6, if both brakes B1 and B2 are open. The same design can be realized also in ranges of axle modules. Braking during free-wheeling needs to be considered appropriately.

Alternatively, free-wheeling can be realized by putting the non-intended axle to torque mode with a small torque setpoint, or to realize better behavior directly with power electronics. This would allow that the main axle can use speed control loop and regulate vehicle speed.

Mechanical decoupling of electric motor would bring benefits of smaller losses in long term free-wheeling, but keeping it coupled would bring benefits in regenerative braking.

### *6.9. Thermal management and redundancy*

Thermal management is a notable differentiator between the two module concepts. In the axle module design, heat sources are concentrated in the central housing containing the electric motor, power electronics and differential. For low-power units, passive cooling through the housing surface may be sufficient; for high-power applications, a single compact cooling circuit can serve the entire axle, simplifying the cooling system design and reducing the number of coolant connections. In contrast, wheel modules distribute the heat sources across four independent locations, each containing its own electric motor and power electronics. While this distributed arrangement reduces the peak heat density at any single point, it necessitates either four independent cooling circuits or an extended central circuit with longer fluid paths, adding complexity and potential points of failure.

Redundancy and fault tolerance differ significantly between the concepts. The wheel module design inherently provides a higher degree of redundancy: with four independent drive units, failure of one module still leaves three operational modules capable of providing propulsion, enabling a limp-home capability. In the axle module design, failure of one of the two modules results in loss of drive to an entire axle, which is a more severe degradation. However, the axle module design benefits from fewer total components (two modules versus four), which statistically reduces the probability of any single failure occurring. The trade-off between higher redundancy with more components (wheel modules) and lower failure probability with fewer components (axle modules) must be evaluated in the context of the specific application requirements.

## 7. RANGES

Both module concepts discussed in this paper require multiple gear ratios between the electric motor and the driving wheels in many cases. However, for small autonomous field robots being transported to the field by a separate vehicle (e.g. low loader), road speed capability may not be required and a single fixed reduction may be sufficient.

The following discussion focuses on full-sized self-propelled machinery. Even in the first successful agricultural vehicles with hydrostatic transmissions, two ranges were needed, one for field and one for road (tractor 656 of International Harvester in 1967) [1]. This was required to a) provide peak efficiency at the operational point for typical heavy field work and b) to provide sufficient efficiency for road speeds.

With electric motors, the number of ranges required depends on the electric motor type used. But it is difficult to achieve high efficiency across the full working speed range without ranges. The required number of ranges depends on the electric motor characteristics and the maximum road speed (typically 25–40 km/h for self-propelled agricultural machinery). The optimal number of ranges is considered to be 2 or more. To simplify calculations, two ranges are discussed in this paper: first for field work (maximum approximately 12 km/h) and second for road drive (maximum 40 km/h).

Let us consider the ground drive of a large self-propelled forage harvester with 200 kW nominal power for the ground drive and two driven axles.

Assuming 800/65R32 wheels (rolling radius approximately 0.78 m), the required maximum wheel speeds are about 41 RPM (at 12 km/h field speed) and approximately 136 RPM (at 40 km/h road speed).

With a nominal speed of the electric motor of 6,000 RPM per axle module, the required total reductions are 146 and 44 respectively, giving a speed ratio spread of the range transmission of approximately 3.3.

Several structural variants exist, including two-shaft designs, standard planetary sets and special planetary configurations. Among these, the special planetary arrangement described by Renius [1] is of particular interest for electric drivelines in agricultural machinery, as it offers not only two adequate ranges but also high internal reductions in both ranges and allows power shifted brakes to be positioned externally.

Systematic analysis of achievable ratio values with different planetary configurations, specifically for ranges in electric drivelines in self-propelled agricultural machinery, remains rare in the literature and presents an important area for further research.

The selection of electric motor nominal speed also affects the range design. Typical nominal speeds for ground drive electric motors are around 6,000 rpm. Higher nominal speeds are generally more favorable in terms of cost per kW and motor compactness, but require greater total reduction and may result in lower driveline efficiency due to increased gear losses. The trade-off between motor speed, reduction complexity and overall driveline efficiency must therefore be considered jointly with the planetary arrangement when designing the range transmission for both axle modules and wheel modules.

## 8. SUMMARY

**Table 1** indicates the pros and cons of each concept rating from 1 to 5, where 5 is the best value.

For self-propelled agricultural machinery, the process drive (chopping, threshing, harvesting) is typically separate from the ground drive in both module concepts.

***Table 1.*** *Evaluation of wheel module and axle module*

| | Wheel module | Axle module |
|---|---|---|
| Freedom for vehicle design | 5 | 3 |
| Scalability | 5 | 4 |
| Manufacturing costs | 3 | 4 |
| Cooling | 3 | 4 |
| Steerability | 5 | 4 |
| Realization of reduction | 4 | 5 |
| Power electronics costs | 3 | 5 |
| Electric motor costs | 3 | 5 |
| Change/repair of the unit | 5 | 3 |
| Controllability (degrees of freedom) | 5 | 3 |
| Replacement of conventional axles | 2 | 4 |

This separation simplifies the module interface: the axle or wheel modules handle only ground drive and steering, while the process drive is powered independently. Weight distribution must be carefully engineered, as self-propelled harvesters carry heavy process equipment and large on-board bunkers, resulting in total vehicle weights that can exceed 60 tons when loaded. Both module concepts allow the frame designer to position the modules to optimize weight distribution for the specific machine type.

The evaluation in Table 1 reflects these considerations: the axle module scores higher on manufacturing cost, power electronics cost and process equipment support, making it the stronger starting point for heavy self-propelled harvesters and root crop machinery; the wheel module scores higher on design freedom, controllability, redundancy and soil compaction management, and is favored where these attributes are the primary design drivers.

While both module concepts permit the vehicle designer to select a single suitable type for a given application, hybrid configurations combining axle modules and wheel modules within the same vehicle are also an option. Such mixed concepts could enable, for example, three-wheel layouts with one wheel module and one axle module; or various layouts with five wheels. This becomes particularly feasible when the same subsystem supplier provides both axle modules and wheel modules within a common product family, so that interface and control compatibility between the modules is inherent. This example illustrates that the two concepts are not mutually exclusive, and that their combination represents an additional dimension of design freedom for novel vehicle architectures of the future.

## 9. CONCLUSIONS

Direct electric drivelines without power-split open fundamentally new design space for self-propelled agricultural machinery and equipment carriers. This paper introduces and compares two modular concepts, axle modules and wheel modules, both defined as self-contained mechatronic units with integrated power electronics, control intelligence and steering, requiring only a DC power bus and communication interface from the vehicle.

The main advantages of the wheel module are: maximum freedom in frame and suspension design, high redundancy with limp-mode capability, full eight-degree-of-freedom controllability enabling torque vectoring, pull-in-turn and precise Ackermann drive, high reduction ratio in both ranges achievable in compact form, and external brake actuation that may serve also as parking brake. Exchange and repair ability is better as well. The main disadvantages are: higher manufacturing and power electronics costs, motor over-dimensioning due to asymmetric wheel loads (e.g. side-hill harvesting), range shift challenges on road, and additional structural elements needed to carry process equipment loads.

The main advantages of the axle module are: lower manufacturing costs per vehicle, automatic handling of asymmetric wheel loads through the differential, structural rigidity suitable for carrying heavy process equipment, simpler centralized cooling, easier regulatory compliance, and easy replacement of the complete axle module. The main disadvantages are: less design freedom between the wheels (material intake of harvesters) and limited controllability with only two degrees of freedom per axle.

The comparison is relevant for self-propelled agricultural machinery, equipment carriers, propelled trailers and field robots, where the design freedom is an important advantage. Several of these machines employ hydrostatic individual wheel or axle drives. The main barrier to replace hydraulics with electrics is current the higher costs, which must be weighed against considerably higher efficiencies resulting in lower energy and $CO_2$ costs. However, costs for electrics may decrease while those for energy will increase. This trend will improve the chances of electrics over time.